\documentclass[11pt]{article}
\usepackage{epsfig}
\usepackage{palatino}
\usepackage{graphicx}  
\begin{document}

\title{$e^+e^-\to e^+e^-\pi^0\pi^0$ at DA$\Phi$NE}
\author{F.~Nguyen$^a$, F.~Piccinini$^b$ and A.D. Polosa$^c$\\
$^a$Dip. Fisica, Universit\`a ``Roma Tre'', and Sez. INFN, via della Vasca Navale 84, Roma, Italy\\
$^b$INFN, Sezione di Pavia and\\ Dip. Fisica
Nucleare e Teorica , via Bassi 6, Pavia, Italy\\
$^c$INFN, Sezione di Roma, p. A. Moro 2, Roma, Italy}
\maketitle
\begin{abstract}
The production of the $\sigma(500)$
meson in $\gamma\gamma\to\pi^0\pi^0$ is studied.
In particular, the KLOE data collected during the DA$\Phi$NE run at 
$\sqrt{s}=1$~GeV are appropriate to this purpose 
because of the strong reduction of Kaon backgrounds.\newline
{\bf Preprint No.} FNT/T/2006-02\newline
\end{abstract}
\newpage

\section{Introduction}
One of the first attempts to describe nucleon-pion interactions within a
spontaneously broken $SU(2)_L\otimes SU(2)_R$ theory was the linear sigma 
model of Gell-Mann and Levy~\cite{gell}. 
The idea is that of writing a Lagrangian of the $\psi=(p~n)$ fields
introducing chirally invariant terms of
the form $\bar{\psi}_L\Sigma\psi_R$ with $\Sigma$ transforming
linearly under chiral transformations: $\Sigma\to L\Sigma R^\dag$. 
Since the defining $SU(2)$ representation is pseudoreal, one can write
$\Sigma$ in terms of only four real parameters. The
linear sigma model parameterization is indeed
$\Sigma=\sigma{\bf 1}+i{\mathbf \tau}^{a}\pi^a$, where $\pi^a$ are the three
isospin components of the pion and $\sigma$ is a scalar field possibly 
associated to some particle in the spectrum. Upon spontaneous symmetry breaking
the nucleon gets a mass $m_N=F_\pi g_{\pi NN}$ in agreement with
the Goldberger-Treiman relation with $g_A$ fixed to be 1
(whereas the physical value is 1.26).
$F_\pi$ is the constant, with dimension of a mass, appearing 
in the potential of the 
model, the well known Mexican hat potential $V=\lambda/4[(\sigma^2+
\mathbf{\pi}^2)-F_\pi^2]^2$. One can show that $F_\pi$ 
so defined coincides with the pion decay constant.

The drawback is that the artificial $\sigma$ field appears 
to be coupled to pions and nucleons suggesting the existence of
a new particle to be looked for.
The natural process where a $\sigma$ contribution is expected to be important 
is the $\pi\pi\to\pi\pi$ elastic channel.
Unfortunately experimental studies have never provided 
over the years a clear signal for it and the assessment of $\sigma$ has 
become more and more controversial. 
The indication coming from $\pi\pi$ collision studies is that,
if the $\sigma$ can be considered as a resonance at all, it ought to be 
an extremely broad (short lived) state.
Very recently~\cite{leut} 
it has been shown that the $\pi\pi$ scattering amplitude 
contains a pole with the quantum numbers of vacuum, the $\sigma$, with a mass
of $M_\sigma=441^{+16}_{-8}$~MeV and a 
width $\Gamma_\sigma=544^{+25}_{-18}$~MeV.
The $\sigma$ has been looked for also in $D$ decays by the E791 Collaboration
at Fermilab~\cite{e791}.
From the $D\to 3\pi$ Dalitz plot analysis, E791 finds that almost the
$46\%$ of the width is due to $D\to\sigma\pi$ with a 
$M_\sigma=478\pm 23 \pm 17$~MeV and
$\Gamma_\sigma=324\pm 40\pm 21$~MeV. 
BES~\cite{bes} has looked for $\sigma$ in $J/\psi\to\omega\pi^+\pi^-$ 
giving a mass value of $M_\sigma=541\pm 39$~MeV and a width of
$\Gamma_\sigma=252\pm 42$~MeV. For a summary of experimental 
data see~\cite{pdg}.

Trying to extend the linear sigma model from $SU(2)_L\otimes SU(2)_R$
to $SU(3)_L\otimes SU(3)_R$ in order to include the strange sector 
($\psi=(u~d~s)$), one encounters the problem that the
fundamental representation of $SU(3)$ is complex and 
a parameterization for the $\Sigma$ field having the same form
of the one given above requires 18 real parameters.
One possibility is to construct
a generalized sigma model with 9 scalar and 9 pseudoscalar fields~\cite{schec}. 
Diverse solutions of this kind have been investigated in the literature but
none of them has proved to be effective in explaining data.

The most successful
approach to build a theory of pions at low energies is that of
Callan-Coleman-Wess-Zumino (CCWZ) 
to define $\Sigma=\exp(2iT^a\pi^a/F_\pi)$ in terms
of the 8 Goldstones arising from $SU(3)_L\otimes SU(3)_R\to SU(3)^{\rm diag}$.
With this definition of $\Sigma$, 
pions do not transform linearly under $\Sigma\to L\Sigma R^\dag$. 
The non-linear realization (one of the infinite realizations, all equivalent
at the level of physical results --the CCWZ theorem)
of chiral symmetry has proven to be extremely successful in the building 
of Chiral Perturbation Theory (ChPT)~\cite{ecker}, 
the standard effective approach
to describe pion interactions at low energies.

The non-linear sigma model excludes the 
$\sigma$ field by construction, therefore the role of a $\sigma$ particle in
$\pi\pi$ interactions, as well as in heavy-light meson decays, calls 
for understanding. Above all, is this a real particle associated 
to some field in an effective Lagrangian (an exotic states Lagrangian 
for example) or is it just a pion 
rescattering effect?

The problem of assessing the existence and nature of this state is not
confined to low energy phenomenology. Just to mention a possible relevant 
physical scenario in which $\sigma$ could play a role,
consider the contamination of 
$B\to\sigma\pi$ in $B\to\rho\pi$ decays (possible because of the large 
$\sigma$ width). This could sensibly affect the isospin analysis for 
the CKM-$\alpha$ angle extraction~\cite{bsig}. Similarly recent studies of the
$\gamma$ angle through a Dalitz analysis of neutral $D$ decays 
need the presence of a $\sigma$ resonance in the fit~\cite{babsig}.

In this paper we want to highlight the possibility that $\sigma$ could be
found in $e^+e^-$ collisions at DA$\Phi$NE now running at a center of
mass energy of 1~GeV, a region where the $\phi$ 
backgrounds are considerably diminished. In particular we examine
the $e^+e^-\to e^+ e^-\pi^0\pi^0$, $\gamma$-fusion channel.
This could represent the cleanest technique available for an independent measure
of the $\sigma$.

\section{The $\gamma\gamma\to\pi^0\pi^0$ channel.}

Our aim is to suggest to extract a $\sigma$ signal in $e^+e^-\to 
e^+e^-\pi^0\pi^0$ 
where also ChPT predictions are very solid:
the $\gamma\gamma\to\pi^0\pi^0$ channel has been determined to
two loop accuracy in the region of photon-photon c.o.m. energy from
about $2 m_\pi$ up to $700$~MeV~\cite{sainio}. In the same energy region the 
$\gamma\gamma\to\pi^+\pi^-$ is affected by a large background $\gamma\gamma\to\mu^+\mu^-$.

To perform our calculation of this process
we will assume that the $\sigma$ is not a standard meson, but a 
crypto-exotic state~\cite{scalmai}. In particular we assume that $\sigma$ is 
$[qq][\bar q \bar q]$ where the parentheses indicate a diquark bound
in the ${\bf \bar 3}_c$ attractive color channel~\cite{wil}. 
Assuming a common spatial 
configuration and spin~0 for the diquark we have a ${\bf\bar 3}_f$ configuration
for the diquark and ${\bf\bar 3}_f$ for the antidiquark because of Fermi
statistics. We can therefore expect to have a full nonet of states 
${\bf 3}_f\otimes {\bf \bar 3}_f={\bf 1}_f\oplus {\bf 8}_f$ whose exotic
content is `crypted' (we would equally have such a nonet with a 
standard $q\bar q$ assignation). Such hypothesis about the $\sigma$
structure is particularly appealing for two reasons. {\bf 1.} It allows to
obtain a mass spectrum for the scalar mesons 
($\sigma, f_0(980), a_0(980), \kappa(900)$) which shows an inverted pattern 
with respect to the $q\bar q$ one. The experimental
spectrum for these states, provided the $\kappa$ is confirmed, resembles the
inverted pattern. {\bf 2.}
Considering the $\sigma$ to be a meson with a different body plan,
qualitatively different from pions, allows to formulate an independent 
effective theory for scalar meson dynamics~\cite{scalmai} escaping
the problem of the emergence of $\sigma$ in ChPT.

\begin{figure}[htb]
\begin{center}
\epsfig{
height=3truecm, width=7truecm,
        figure=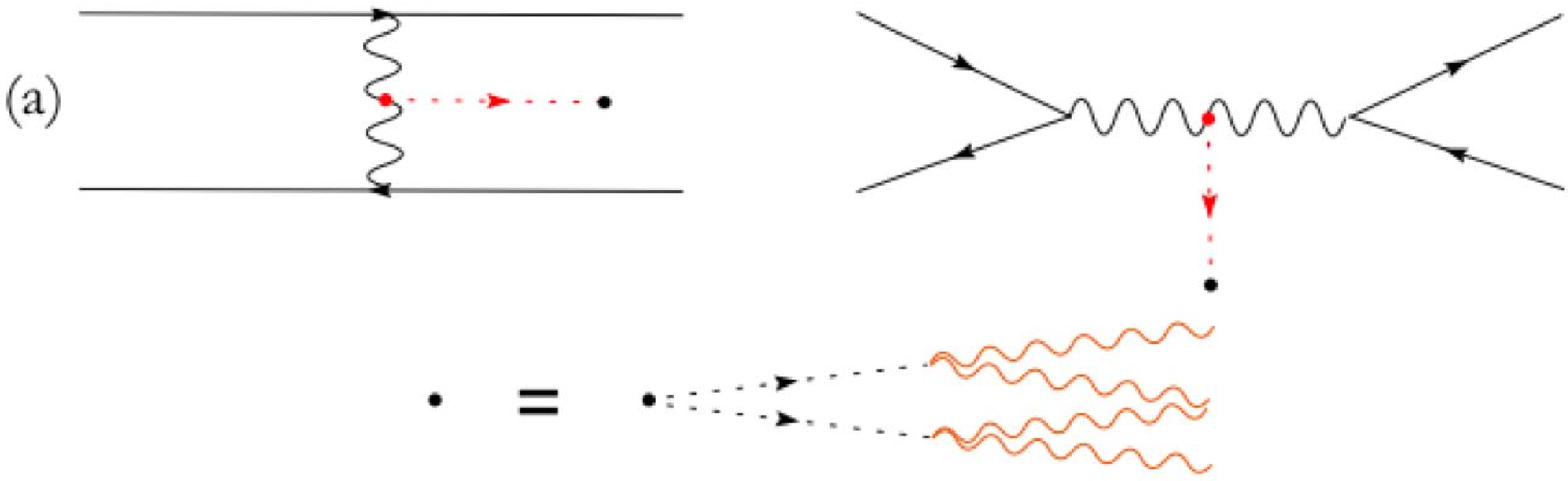}
\hspace{1in}%
\epsfig{
height=3truecm, width=4truecm,
        figure=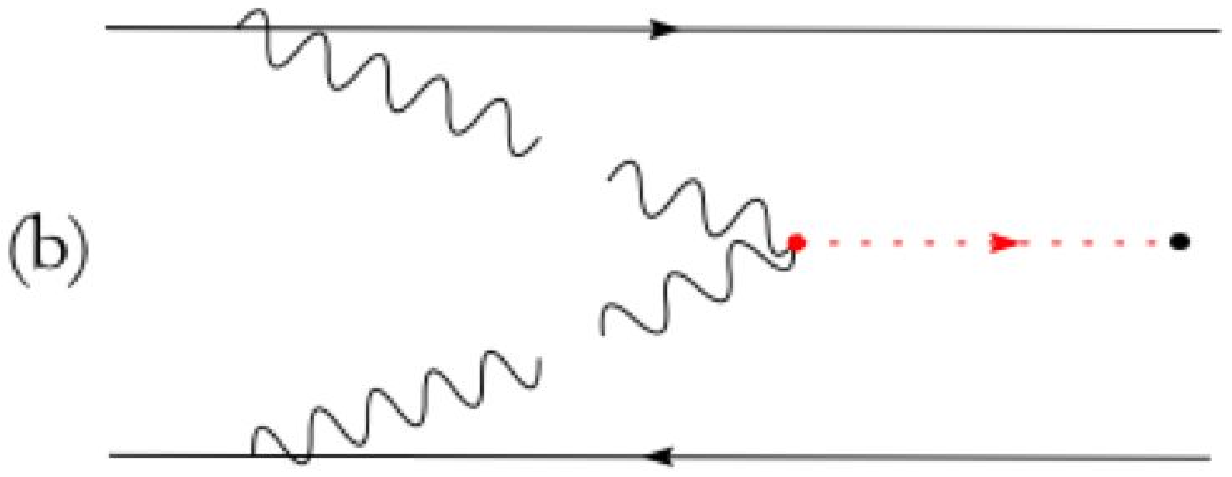}
\caption{\footnotesize 
(a) The Feynman diagrams for $e^+e^-\to e^+e^-\pi^0\pi^0$. 
In the $t$-channel one photon is exchanged between electron and positron
and a $\sigma$ is coupled to it. The black dot indicates the
decay of $\sigma\to\pi^0\pi^0$ where the two pions are allowed to decay
$\pi^0\to\gamma\gamma$. The $\gamma$'s are eventually measured
in the detector. The final-particle phase space in our calculation
is the 4-body $e^+e^-\pi^0\pi^0$. In the $s$-channel a virtual photon emits a 
$\sigma$ and another virtual photon, eventually decaying into $e^+e^-$.
(b) The $\sigma(\gamma\gamma\to\pi^0\pi^0)$ cross section is factorized 
with the distribution functions of the photon in the electron. Such
approximation is strictly valid for collinear photons, being less and less
applicable once one allows for a $p_\perp\neq 0$ of the electron.
}
\label{f:diag}
\end{center}
\end{figure}

Our calculation will be performed using the full matrix element derived 
by the Feynman amplitudes in Fig.~\ref{f:diag}(a). To further test 
our results, we also adopt the double equivalent photon approximation,
see Fig.~\ref{f:diag}(b), largely used in the past for studying this
kind of processes~\cite{brodsky,panca}.

\begin{figure}[htb]
\begin{center}
\epsfig{
height=7truecm, width=10truecm,
        figure=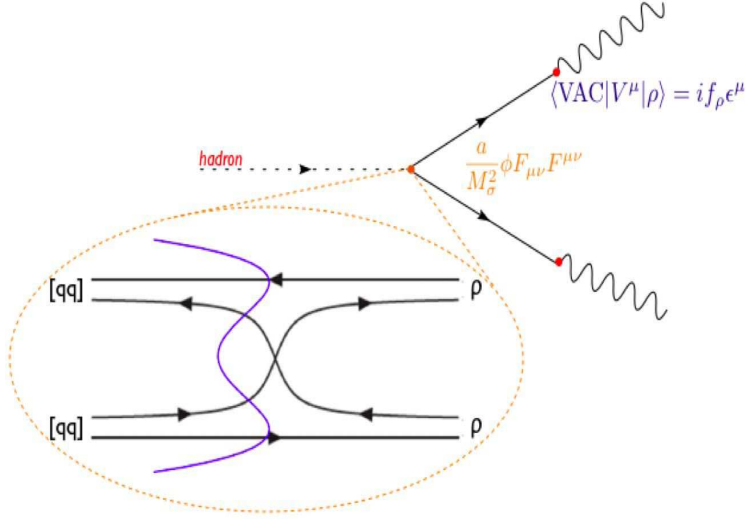}
\caption{\footnotesize 
The vertex $\sigma\gamma\gamma$ is obtained assuming 
Vector-Meson-Dominance (VMD): the $\sigma$ decays
to $\rho\rho$ and each $\rho$ converts to $\gamma$ according to its $f_\rho$
decay constant. Our microscopic picture of the decay process of 
a $\sigma$ to $\rho\rho$ is a kind of
tunneling of a quark which escapes its diquark shell to meet an antiquark
from the antidiquark forming a standard color singlet $q\bar q$ meson.
The higher the barrier to be crossed the stronger is the diquark energy
binding.
The interaction strength $a$ is expressed as the coupling of a 
gauge-invariant term $a/M_\sigma^2 \phi F^{\mu\nu}F_{\mu\nu}$, $\phi$
being the field associated to the $\sigma$~\cite{scalmai}.
}
\label{f:vert}
\end{center}
\end{figure}

\begin{figure}[htb]
\begin{center}
\epsfig{
height=7truecm, width=10truecm,
        figure=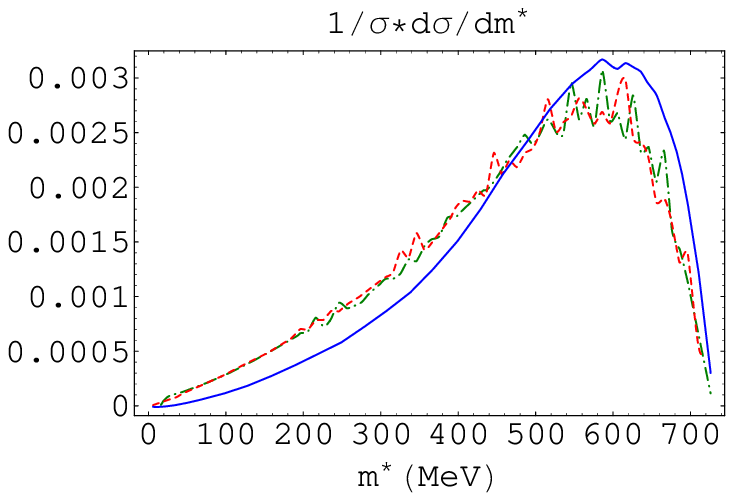}
\caption{\footnotesize 
The solid line represents the so
called double equivalent photon approximation; see Fig.~1(b)~\cite{panca}.
Dashed and dot-dashed represent the $t$ and $s+t$ channel respectively
in the Feynman diagram computation of Fig.~1(a).
By $m^*$ we denote the invariant mass of the $e^+e^-$ final state pair 
(missing mass). With a moderate statistics simulation, the four-body phase 
space of the matrix element calculation converges with more difficulty
with respect to the two body phase space of the double equivalent 
photon approximation.
}
\label{f:mmass}
\end{center}
\end{figure}

The double equivalent photon approximation is strictly valid when the 
photons emitted by the electrons (see Fig.~\ref{f:diag}(b)) are collinear.
On the other hand the results of our, straightforward, 
calculation are formally valid also
for final state electrons selected with a certain $p_\perp$. In such a way
one should be able, at least in principle, to perform
the experimental analysis also in absence of forward 
electron detectors (electron tagging). 

The model dependent part of the calculation is in the $\gamma\gamma\sigma$
vertex. In Fig.~\ref{f:vert} we show how we do parameterize such interaction.
Because of gauge invariance, the vertex has the form $a/M_\sigma^2 \phi
F^{\mu\nu}F_{\mu\nu}$, $\phi$ representing the $\sigma$ field. In a previous
paper~\cite{scalmai} the method for the computation of the coupling of
a cryptoexotic $\sigma$ to $\rho$ mesons is traced. The coupling $a$ used
here needs not to be equivalent to the ${\cal A}=2.6$~GeV coupling fitted 
in~\cite{scalmai}. ${\cal A}$ defines the decay vertex $\sigma\to\pi\pi$. 
Since we cannot fit $a$ from data we treat it as a free
parameter. This doesn't affect our analysis since we will only discuss
normalized distributions.  In other words the 4-quark 
hypothesis does not alter quantitatively the nature of our conclusions
though it is intimately connected to the qualitative picture of the 
microscopic dynamics of the $\sigma \to \rho\rho$ decay.
The $\rho$ mesons are coupled to photons using a VMD Ansatz, 
see Fig.~\ref{f:vert}.

The $\sigma$ propagation is described by a simple Breit-Wigner function
with mass and width taken from E791 data.
The cross section can be plotted as a function of several variables.
We choose to show the distributions in the missing mass $m^{*}$, 
i.e., the invariant
mass of the final state $e^+e^-$ pair and in terms of 
the invariant mass of the incoming $\gamma\gamma$, $W_{\gamma\gamma}$.

\section{Results}
The reason for plotting the cross section distribution as a function
of the missing mass $m^*$ of the final $e^+e^-$ pair, Fig.~\ref{f:mmass} is that using
this variable we can get rid of the most relevant background we have,
namely $e^+e^-\to\omega\pi^0\to \pi^0\pi^0\gamma$. This is peaked around
zero in the $m^*$ variable and the distribution smearing could still 
allow the
signal from background extraction.

\begin{figure}[htb]
\begin{center}
\epsfig{
height=7truecm, width=10truecm,
        figure=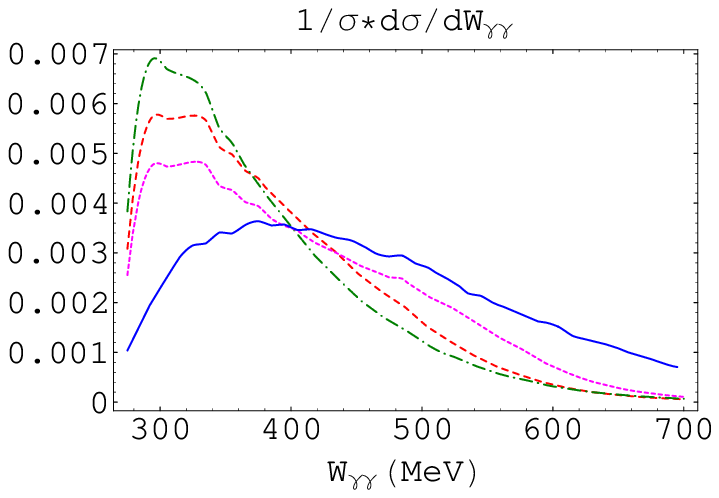}
\caption{\footnotesize 
$W_{\gamma\gamma}$ is the photon-photon center of mass energy. The solid line
is the $e^+e^-\to e^+e^-\pi^0\pi^0$ cross section distribution calculated
convoluting the two-loop $\sigma (\gamma\gamma\to\pi^0\pi^0)$ cross
section with the photon distribution functions according to the
double equivalent photon approximation. Dashed, dot-dashed and dotted lines
represent the same quantity 
where the two pions come from the decay of a $\sigma$ produced in photon-photon
fusion. In particular the dashed curve is computed using the $\sigma$ parameters
found by E791 ($M_\sigma=478$~MeV and $\Gamma_\sigma=324$~MeV)~\cite{e791}, 
the dot-dashed corresponds to an $M_\sigma=441$~MeV and
$\Gamma_\sigma=544$~MeV (from the $\pi\pi$ scattering amplitude 
analysis~\cite{leut}) while the dotted curve is computed
according to BES  results ($M_\sigma=541$~MeV and $\Gamma_\sigma=252$~MeV)~\cite{bes}. 
}
\label{f:chpt}
\end{center}
\end{figure}

$\gamma\gamma\to\pi^0\pi^0$ has been computed in ChPT
to two-loop level accuracy~\cite{sainio}. We can as well plot the normalized distribution
$1/\sigma*d\sigma/dW_{\gamma\gamma}$, $W_{\gamma\gamma}$ being 
the photon-photon center of mass energy, see Fig.~\ref{f:chpt}.
The two-loop ChPT cross section $\sigma(\gamma\gamma\to
\pi^0\pi^0)$ is convoluted with photon distribution functions as prescribed by
the double equivalent photon approximation.

It is not surprising that the effect of a Breit-Wigner resonance 
makes this distribution different from what expected in 
ChPT where the $\sigma$ can only be simulated by an effect of strong
interaction in the $\pi\pi$ channel.
Assuming that $\sigma$ is a propagating particle, with either 4-quark or 
2-quark structure, qualitatively changes the $d\sigma/dW_{\gamma\gamma}$
with respect to ChPT predictions.
Anyway the $W_{\gamma\gamma}$ region where our calculation differs more
sensibly from ChPT could be affected by interference effects with $\gamma\gamma\to f_0(980)\to \pi^0\pi^0$. However to address this issue will require precise data at much higher $\sqrt{s}$ values.

\begin{figure}[htb]
\begin{center}
\epsfig{
height=6.5truecm, width=9truecm,
        figure=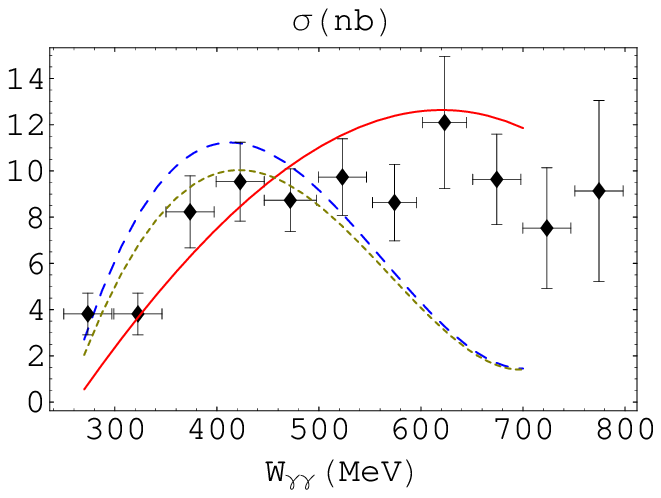}
\caption{\footnotesize 
Data points are cross section values for $\gamma\gamma\to\pi^0\pi^0$ obtained by Crystal Ball 
(1993). The solid curve represent the two-loop ChPT result . The dashed line
is $\sigma(\gamma\gamma\to\sigma\to\pi^0\pi^0)_{\rm B.W.}$. 
The dotted line is the same as the dashed re-weighting the amplitude by an Adler zero 
factor~\cite{adler} $(s-s_A)$ where $s_A=0.5m_\pi^2$. 
Data uncertainties
certainly call for a more precise measurement of this channel.
}
\label{f:combi}
\end{center}
\end{figure}

In Fig.~\ref{f:combi} we compare our results 
to the only existing data set for the $\sigma(\gamma\gamma\to\pi^0\pi^0)$, namely the 1993
Crystal Ball data points. The solid line represent the two-loop ChPT result . The dashed line
is $\sigma(\gamma\gamma\to\sigma\to\pi^0\pi^0)_{\rm B.W.}$ where the $a$ parameter has been fitted
to the cross section value of the lowest $W_{\gamma\gamma}$ experimental point.
The dotted line is the same as the dashed re-weighting the amplitude by an Adler zero 
factor~\cite{adler} $(s-s_A)$ where $s_A=0.5m_\pi^2$. 
The resonant contribution should be properly combined with the chiral loop amplitudes to show
a more realistic result: the two contributions do not exclude each other. 

Our results are certainly dependent on the Breit-Wigner (BW) parameterization of the $\sigma$ 
contribution (we have also analyzed the possibility of implementing the BW Ansatz with a 
$\sigma$ comoving width).
However, given the poor resolution expected for  low energy photons and the actual absence of an electron tagging device, we believe that our simple approach is appropriate enough to motivate 
an experimental analysis in this channel: Fig.~\ref{f:combi} certainly enforces the case to require more precise data. The use of a Breit-Wigner parameterization of $\sigma$ at E791 and BES is probably more challenged by the fact that charged pions momenta are more precisely measured in these experiments, allowing a quite good resolution on invariant masses.
Moreover the recent experience at BaBar~\cite{cavoto} where the $D(D_s)\to 3\pi$ Dalitz plots are currently under study, has shown preliminarly that the use of more sophisticated approaches  to the parameterization of the isoscalars (e.g. K-Matrix) are not providing more stringent and convincing results in the extraction of the $D\to \sigma\pi $ channel with respect to what obtainable via simple BW Ansatz.

\section{Backgrounds}

\begin{figure}[htb]
\begin{center}
\epsfig{
height=8truecm, width=8truecm,
        figure=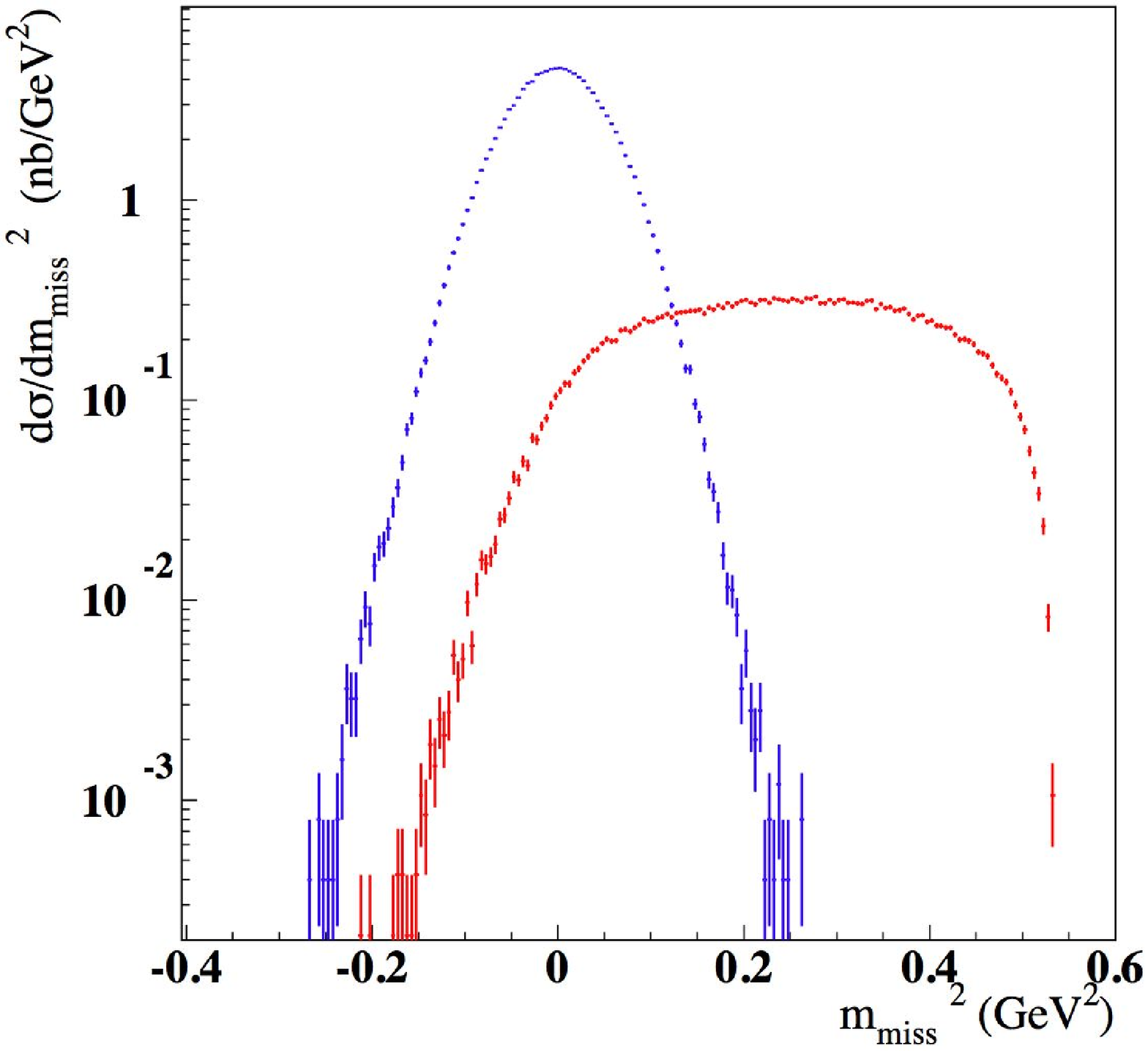}
\hspace{0.001in}%
\epsfig{
height=8truecm, width=8truecm,
        figure=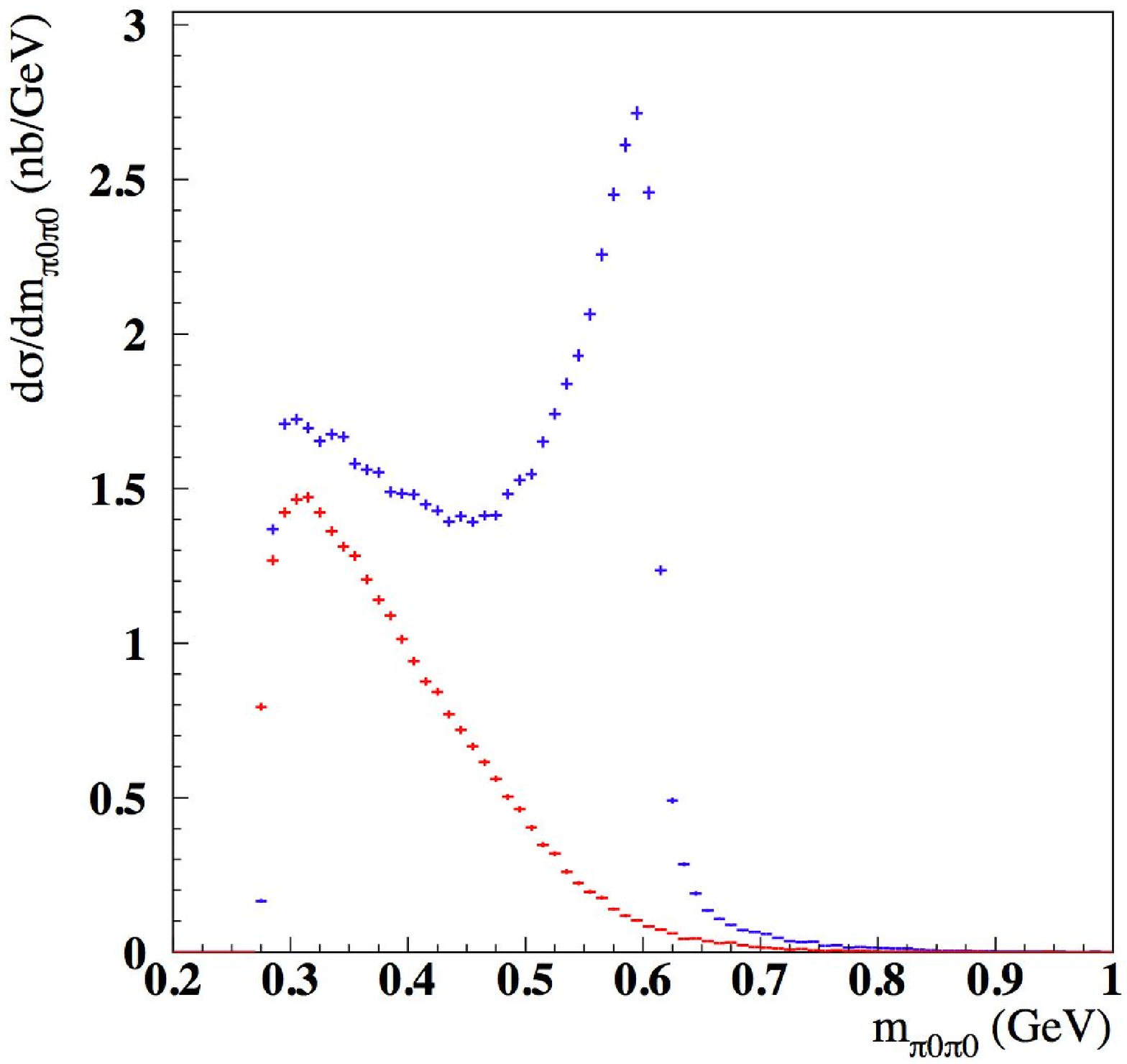}
\vspace{-1.5truecm}        
\caption{\footnotesize 
In the left panel we compare the background $e^+e^-\to \omega\pi^0$ 
distribution (the one centered around zero) to the 
signal (the broader distribution)  as functions of
the missing mass $m_{\rm miss}\equiv m^*$. In the background we have five photons
in the final state. We pick up four of them reconstructing 
$\pi^0\pi^0\to 4\gamma$ and perform the smearing, as explained in the
text.
Even after smearing the signal and the $\omega\pi^0$ 
background appear to have different shapes in $m^*$.
In the right panel we perform the same comparison using another variable, 
namely, the invariant mass of the $\pi^0\pi^0$ system. The peak is in 
correspondence of two back-to-back $\pi^0$'s; the shape at $\sim 280$~MeV
is the $2 m_\pi$ invariant mass.
}
\label{f:smear}
\end{center}
\end{figure}

Here we proceed to a list of the relevant backgrounds which have to be 
taken under control in the data analysis. In fact, even if KLOE
is collecting data at $\sqrt{s}=1$~GeV, some of the $\phi$ decays
with at least 4 photons in the final state have a rate comparable
to the signal one (we require at least 
$4\gamma$'s in the final state).
 
\begin{itemize}
\item $\phi\to\eta\gamma\to 3\pi^0\gamma.$ At a center of mass energy of 
$1$~GeV the $\phi$ is strongly reduced. The cross section 
for $e^+e^-\to\phi\to\eta\gamma$ drops by about a factor of 
30~\cite{Akhmetshin:2004gw}
from $~20$~nb at $\sqrt{s}=M_\phi$ down to $0.67$~nb 
at $\sqrt{s}\simeq1$~GeV;
since the branching ratio $BR(\eta\to 3\pi^0)\simeq 32.5\%$,
we have $\sigma(e^+e^-\to\phi\to\eta\gamma\to 3\pi^0\gamma)\sim 0.2$~nb,
at $\sqrt{s}\simeq1$~GeV.
\item $\phi\to K_S K_L.$ The cross section in this case drops from
$1350$~nb to $11$~nb according to the indications 
of CMD-2~\cite{Akhmetshin:1999ym}.
This background is particularly dangerous when $K_S$ decays into $2\pi^0$'s
($BR(K_S\to 2\pi^0)\simeq31\%$) and $K_L$ escapes detection.
Considering that in KLOE at $\sqrt{s}=1$~GeV, about $75\%$ of $K_L$'s decay inside the detector,
and that about $70\%$ of the surviving ones interact in the Electromagnetic Calorimeter
we get a further reduction of the $\sigma(e^+e^-\to\phi\to
K_S K_L\to\pi^0\pi^0 K_L)$ (with $K_L$ undetected) 
which amounts to $~0.2$~nb.
\item $\phi\to\pi^0\pi^0\gamma.$ Starting from the
KLOE result~\cite{Aloisio:2002bt}
$BR(\phi\to f_0(980)\gamma)\simeq 10^{-4}$ at $\sqrt{s}=M_\phi$, 
and assuming that $\sigma(e^+e^-\to\phi\to f_0(980)\gamma)$
gets reduced by a factor 30 at $\sqrt{s}=1$~GeV, 
we estimate a cross section of about 
$\sim 10$~pb for this channel.
\item $\phi\to \pi^0\eta\gamma.$ Given the KLOE result~\cite{Aloisio:2002bs}
$BR(\phi\to a_0(980)\gamma)\simeq 7\times10^{-5}$ at $\sqrt{s}=M_\phi$,
and taking into account $BR(\eta\to\gamma\gamma)\simeq 39.4\%$ and the
reduction by a factor of 30, we estimate a cross section of about
$\sim 4$~pb for this channel.
\item $e^+e^-\to\eta e^+e^-\to \pi^0\pi^0\pi^0 e^+e^-.$ 
We estimate a cross section for this process amounting to $\sigma\simeq 13$~pb.
Such a background should be removable imposing that the 4
reconstructed photons belong to 2 pions going
back-to-back in the transverse plane.
\end{itemize}

\subsection{The $e^+e^-\to \omega\pi^0$ channel}
Contrary to $\phi$ decays, the cross section for
this process tends to increase from $\sigma=(0.51\pm0.07)$~nb 
at $\sqrt{s}=1.02$~GeV to $\sigma=(0.56\pm 0.14)$~nb 
at $\sqrt{s}=1.005$~GeV, as measured by the
SND~\cite{Achasov:1999wr} experiment. Moreover the peak at $~0.6$~GeV in the $\pi\pi$
invariant mass, as shown in the right panel of Fig.~6, corresponds to events with 2 pions 
emerging back-to-back in the transverse plane.

Since the fraction of background events in which the fifth photon is lost has the same experimental
signature of the signal, namely $2\pi^0$'s collinear in the transverse plane plus missing energy,
we performed a dedicated study of this channel.
We consider $e^+e^-\to \omega\pi^0\to\pi^0\pi^0\gamma$ 
to proceed via VMD $e^+e^-\to\gamma^*\to\rho\to\omega\pi^0$.
This channel has a missing mass peaked at zero.
However, the energy resolution of the photons coming from the 
$2\pi^0$'s modifies this simple pattern.

Every photon energy distribution is convoluted
with a Gaussian function, where the standard deviation
is the energy resolution function of the KLOE
Electromagnetic Calorimeter~\cite{Adinolfi:2002zx}:
$$
\frac{\sigma_E}{E} ~ ~=~ ~ \frac{5.7\%}{\sqrt{E~[\mathrm{GeV}]}}.
$$
The same procedure is applied to the 4 photons
generated by the signal $e^+e^-\to\pi^0\pi^0 e^+e^-$.

Upon such a smearing procedure, no significant
change is introduced in the signal distribution,
while the $\delta$-function in
the missing mass gets broadened but still does not
superimpose to the 
signal region, see Fig.~\ref{f:mmass} and Fig.~\ref{f:smear}
where the cross section value
$\sigma(e^+e^-\to\omega\pi^0\to\pi^0\pi^0\gamma)=0.6$~nb is used. 

It is also clear that the experimental analysis is going to
deal not only with the photon energy resolution, but
also with the photon efficiency as a function
of the energy and of the polar angle
given the correct pairing
of the final photons to the parent $\pi^0$'s. A full
simulation of the detector would be in order for a detailed 
discussion of these issues.
Nevertheless, it is reasonable to guess that the same
sources of inefficiency are shared by the
signal and the background processes.

\section{Conclusions}
We believe that KLOE in the low energy DA$\Phi$NE 
run has a concrete opportunity to find (or disprove) the
$\sigma$ in a clean experimental channel, $\gamma\gamma\to\pi^0\pi^0$.
The scope of our feasibility study is to underscore this possibility and to
suggest some data analysis strategies. The crucial interest should be 
that of investigating the nature of the $\sigma$: is it a particle or
an effect of chiral dynamics? 
The comparison with ChPT predictions for 
$\gamma\gamma\to\pi^0\pi^0$ illustrates that if a broad resonance 
is indeed produced in $\gamma\gamma$ it could be detectable in the 
low energy $\gamma\gamma$ region, see Fig.~\ref{f:chpt},
from the distribution slopes.
As soon as smearing is introduced the two curves (resonance-type versus ChPT) will tend to
superimpose. This certainly calls for a more selective analysis of data
making use of forward detectors to tag electrons. Such devices, which at the moment
are not part of the experimental apparatus, could be integrated in view of possible future luminosity upgrades.

\section*{Acknowledgements}
We wish to thank F.~Ambrosino,  D.~Babusci, C.~Bloise, F.~Ceradini, G.~Capon,
G. D'Ambrosio,
S.~Giovannella, M.~Sainio and G.~Pancheri 
for useful discussions. In particular we wish to
thank L.~Maiani for his comments on the manuscript.

\end{document}